\documentstyle[preprint,revtex,eqsecnum]{aps}
\begin{document}
\draft
\begin{title}
\bf{DIRECTED POLYMERS WITH \\ RANDOM INTERACTION :\\
AN EXACTLY SOLVABLE CASE}
\end{title}
\author{Sutapa Mukherji\cite{eml1}
and Somendra M. Bhattacharjee\cite{eml2}}
\begin{instit}
Institute Of Physics, Bhubaneswar 751 005, India
\end{instit}
\begin{abstract}
We propose a model for two $(d+1)$-dimensional directed
polymers subjected to a mutual $\delta$-function
interaction with a random coupling constant, and present an
exact renormalization group study for this system.  The
exact $\beta$-function, evaluated through an
$\epsilon(=1-d)$ expansion for second and third moments of
the partition function, exhibits the marginal relevance of
the disorder at $d=1$, and the presence of a phase
transition from a weak to strong disorder regime for $d>1$.
The lengthscale exponent for the critical point is
$\nu=1/2\mid\epsilon\mid$. We give details of the
renormalization.  We show that higher moments do not require any
new interaction, and hence the $\beta$ function remains the
same for  all moments.  The method is extended to
multicritical systems involving an $m$ chain interaction.
The corresponding disorder induced phase transition for
$d>d_m=1/(m-1)$ has the critical exponent
${\nu}_m=[2d(m-1)-2]^{-1}$. For both the cases, an
essential singularity appears for the lengthscale right at
the upper critical dimension $d_m$.  We also discuss the
strange behavior of an annealed system with more than two
chains with pairwise random interactions among each other.

\end{abstract}
\narrowtext

\section{INTRODUCTION}
\label{sec:intro}

Attempts to study the effects of randomness, especially if
one requires averages of thermodynamic quantities, have led
to many new techniques, concepts, and, probably,
controversies.\cite{mpv} In order to get a clear idea about
random systems, in recent years, a directed polymer (DP) in
a random medium seems to have emerged as the consensus
candidate for the ``simplest" random model.
\cite{karnel,kar,der90,par90,kim} We here
propose a still simpler problem of DPs with random
interaction that can be solved using an exact field
theoretic renormalization group (RG) approach. \cite{smbsm}
This, we believe, is highly significant since RG is the
general framework to study and to understand, through the
fixed point spectrum, the universal aspects of any model.

Directed polymers in $(d+1)$-dimensions are random walks
directed along a particular direction, say z, with
fluctuations in the transverse $d$-dimensional space.  DPs
are of considerable interest and have attracted a lot of
attention as a simple statistical mechanical model because
of its relevance and applicability in unifying a wide
variety of seemingly disparate systems.  These include the
flux lattice melting problem in high $T_c$
superconductors,\cite{nel88} commensurate-incommensurate
transitions,\cite{fish84} wetting
transition,\cite{prvs,forg} vertex models,\cite{smbjj}
polymeric nematics, \cite{polnem} biomembrane phase
transitions,\cite{ia} interface growth,\cite{kpz} etc.
Many problems of conventional polymers [self avoiding walks
(SAW)] like collapse, adsorption etc, have exactly solvable
counterpart in DPs.\cite{yeo} The RG analysis of a pure
system of interacting DPs gives enough insights through the
evaluation of the exact $\beta$-function to all orders in
perturbation series.\cite{jjb,smbjj,smbph} These systems of
DPs with pure short range interactions are almost
completely solved, and, for example, have led to several
exact results for vertex models.\cite{smbjj,smbjj2}

There are many efforts and activities in the field of
polymers with random interaction \cite{polint} or in random
media\cite{polmed}. The analogous DP problems are expected
to be simpler. For example, a DP in a random medium, which
through a nonlinear mapping describes many aspects of
interface growth, has been studied upto one loop in the
momentum shell technique.\cite{kpz} There were also
attempts for solving the many chain system in a random
medium in the context of high $T_c$ superconductivity.
\cite{nat} It is the directedness that helps
in setting up the DP problem, both analytically and
numerically, as oppposed to the SAW problem in random
media.  Several results for the DP problem are known in
general, though exact or rigorous results are rather
few.\cite{kar} Apart from these random media problems, the
other category of problems involve polymers with random
interactions in the context of, say, disordered
heteropolymers.\cite{polint} Here again, a DP with random
interaction turns out to be simpler.\cite{dgh}

Our model\cite{smbsm} has similarity with the second
category of problems. It deals with a random {\it mutual}
interaction among the chains with the randomness in the
coupling constant of the interaction. The randomness is
only along the length of the chains and does not depend on
the transverse $d$ dimensional coordinates. The specific
charateristic of the randomness as well as the directed
nature of the polymers enable us to solve the model
exactly.  We, furthermore, show that this model, inspite of
its simplicity, captures many of the essential features
such as marginal relevance, existence of a disorder induced
phase transition, etc. as known, e.g., for the interface
growth problem, DP and SAW in random environments
etc.\cite{der90,polmed} The two dimensional wetting
phenomenon is also analogous to our proposed system
\cite{forg,derhv} - though our model (and the solution) is for
general $d$.

We define the model in the next section, and to put things
in the proper context, the aim and the outline of the paper
are given there.
\section{MODEL}
\label{sec:model}
{}From the definition of DPs, it follows that a projection of
a DP in the transverse $d$-dimensional space is an ordinary
polymer with $z$ representing the contour variable which is
equivalent to the steplength in a discrete case.  In the
path integral formulation the dimensionless hamiltonian for
two such DPs, each of length $N$, interacting through a
random mutual shortrange interaction can be written as
\cite{smbsm}
\begin{equation}
{\sf H}=\frac{1}{2} \int_{0}^{N} dz \
\left[\left(\frac{\partial
{\bf{r}}_1(z)}{\partial z}\right)^{^{\scriptstyle 2}}+
\left(\frac{\partial{\bf{r}}_2(z)}{\partial
z}\right)^{^{\scriptstyle 2}}\right]+
\int_{0}^{N} dz \ v_0\ [1+b(z)]\ V{\bf
(}{\bf{r}}_{12}(z){\bf )}.
\label{eq:h}
\end{equation}
where ${\bf r}_i(z)$ is the $d$-dimensional position vector
of a point of chain $i$ at a contour length $z$, and ${\bf
r}_{12}(z)={\bf r}_1(z)-{\bf r}_2(z)$. The first term that
comes from the chain connectivity is the entropic
contribution, and corresponds to free chains.\cite{path}
The second term is the two chain interaction at the same
chainlength through a short range potential $V({\bf{r}})$.
We introduce the randomness through the coupling constant.
It has a pure part $v_0$ and a random part $v_0b(z)$ which
varies only with $z$ (the length along the chain).  It is
chosen in this way so that $b(z)$ is dimensionless.  At
this stage, for generality, we keep $V$ as a short range
potential.  Later on, specific calculations would be done
with a $\delta$-function potential.  Also, starting with a
short range potential has certain mathematical advantages
like avoiding powers of distributions, as we will see
below.  One can also think of this problem as a
nonrelativistic quantum problem of particles with time
$(z)$ dependent interaction potential - a description we do
not find very illuminating.

One of the simplest but nontrivial choices for the
distribution of the randomness is a Gaussian one:
\FL
\begin{mathletters}
\begin{eqnarray}
P{\bf (}b(z){\bf )}=(2\pi\Delta)^{-1/2}\
\exp[-b(z)^2/(2\Delta)],
\label{eq:disa} \\
\langle b(z)\rangle=0, \ {\rm and}\ \langle b(z_1)
b(z_2)\rangle\ =\Delta\
\delta(z_1-z_2).\label{eq:disb}
\end{eqnarray}
\end{mathletters}
Here the randomness is uncorrelated in nature and is
described by the variance $\Delta$.  Choosing a zero mean
for $b(z)$ is not a restriction because any nonzero $\big
<b(z)\rangle$ could be ``gauged away" by absorbing it in
the pure part.

So far we have discussed only two body interactions. For
DPs it is known that even pure many body interactions,
representing special multicritical points, can also be
handled exactly. \cite{jjth,smbjj2,smbph} It is possible to
study the disordered versions of these multicritical
systems. The hamiltonian for the $m$-th order multicritical
point, involving only $m$-body $\delta$-function
interaction, is
\begin{equation}
{\sf H}_m=\frac{1}{2} \int_{0}^{N} dz \
\sum_{i=1}^{m}\left(\frac{\partial
{\bf{r}}_i(z)}{\partial z}\right)^{^{\scriptstyle 2}}+
\int_{0}^{N} dz \ v_m\ [1+b(z)]\ \prod_{i=1}^{m-1}
\delta{\bf(}{\bf
{r}}_{i\ i+1}(z){\bf )}\label{eq:hm}
\end{equation}
where, as before, $b(z)$ is the random part. We come back
to this multicritical situation in section
{}~\ref{sec:multi}.  It is also possible to define more
general systems by putting the lower order interactions in
Eq. ~\ref{eq:hm} with independent random coupling
constants.  Such a hamiltonian can, in principle, describe
the approach to the multicritical points.  However, such
complicated cases are not discussed here.

One possibility of getting a random interaction for, say,
the two chain case, is to take ``charged" DPs, with random
charges $q_i(z)$ for the $i$th chain, $v_0b(z) = q_1(z)
q_2(z)$, and interactions of charges only at same $z$.  If
the charges are in thermal equilibrium with the polymers, a
simple quadratic hamiltonian for the charges can be taken
as proportional to $\int b(z)^2 \ dz$, and it is to be
added to the hamiltonian of Eq. ~\ref{eq:h}.  The partition
function one gets from this full hamiltonian is really
equivalent to $\langle Z\rangle$ (annealed case) evaluated
with the hamiltonian of Eq.  ~\ref{eq:h} but averaged over
the distribution of Eq.  ~\ref{eq:disa}.  The more complex
situation is the quenched average which requires, e. g.,
the average of the free energy.

One of the standard approaches for random systems is to
proceed through the evaluation of the quenched free energy
using the replica trick \cite{mpv}
\[\langle\ln{Z}\rangle=\lim_{n\rightarrow\infty}
\frac{\langle Z^n\rangle-1}{n}\]
which requires evaluation of $\langle Z^n\rangle$.  The
importance of these moments can be realized through the following
expansion
\[\langle\ln{Z}\rangle = \ln\langle Z\rangle+
\sum_{n=2}^{\infty}\ n^{-1}\langle
\big[Z/\langle Z\rangle-1\big]^n\rangle.\]
Such an expansion makes sense if and only if the various
cumulants of the partition function, with respect to the
disorder distribution, do not grow too rapidly with $n$. In such
a case there will not be much qualitative difference between the
quenched and annealed cases. This, in turn, suggests that to
look into the possible differences one can study the various
moments of the partition function. In addition, the moments can
be looked upon as the charateristic function for the probability
distribution of $\ln Z$.\cite{mackir} They are, therefore, of
interest in themselves.\cite{lub} This is the approach we take
in this paper.  Our analysis is not yet enough for the analytic
continuation in $n$ to $0\leq n\leq 1$ regime, as one would need
for the free energy.  This is not a deterrence as important
information can be gathered even from the integral moments.

Instead of evaluating the quenched free energy, we
systematically study the behavior of $\langle Z\rangle$,
the second and the third cumulants.  The first moment
describes the behavior of an annealed system while the
higher cumulants would show the nature of fluctuations. For
each case, we do the averaging exactly before the
configuration sum (or ``path integrals") to define an
effective hamiltonian for that particular cumulant. This
``pure" effective hamiltonian is then treated by
perturbative renormalization. The effect of disorder is
felt through the generation of new terms in the effective
hamiltonian. The RG analysis helps in examining the flow of
these terms as the lengthscale is increased, thereby,
showing the marginal relevance or irrelevance of the
disorder.

In section ~\ref{sec:zav}, $\langle Z\rangle$ is
discussed, while $\langle Z^2\rangle$ and $\langle
Z^3\rangle$ are done in section ~\ref{sec:z2av} and
{}~\ref{sec:z3av}. Though the derivation of effective
hamiltonians preceeds perturbative analysis in sections
{}~\ref{sec:zav} ,~\ref{sec:z2av} and ~\ref{sec:z3av}, it is
instructive to start with the original hamiltonian, do the
perturbation analysis and then do a term by term disorder
averaging. Such a procedure not only shows how the new
terms are generated but also acts as a cross check. This is
discussed in Appendix A. Appendix B discusses many of the
details needed in section ~\ref{sec:z2av}. The random
multicritical case is discussed in section
{}~\ref{sec:multi}. The annealed case of three and four
chains is discussed in section ~\ref{sec:big}. The paper
ends with a discussion and summary in section
{}~\ref{sec:smry}.

\section{$\langle  Z \rangle$}
\label{sec:zav}
We show in this section that the annealed case can be
reduced to a pure problem. We add that this reduction is
special for two chains.  Had we started with more than two
chains, say three or four, with the same random pairwise
interaction as in Eq. ~\ref{eq:h}, the annealed case would
be completely different from the corresponding pure case,
and would have a much richer structure.  A particular case
is discussed in section ~\ref{sec:big}.

The partition function, in the continuum approach, for a
system of two chains, given by the hamiltonian of Eq.
{}~\ref{eq:h}, is
\begin{eqnarray*}
Z=\int Dr_1\ Dr_2\ \exp (-{\sf H})
\end{eqnarray*}
where $\int Dr_1 Dr_2$ stands for the sum over all
configurations of the two chains.  A straightforward
averaging of $Z$ using the probability distribution of
Eq.~\ref{eq:disa} defines an effective hamiltonian ${\cal
H}_{{\rm eff}}$ such that
\begin{equation}
\langle Z\rangle=\int Dr_1\ Dr_2\ \exp
(-{\cal H}_{{\rm eff}}),\label{eq:za}
\end{equation}
and it is given by
\begin{equation}
{\cal H}_{{\rm eff}}=\frac{1}{2} \int_{0}^{N} dz \
\sum_{i=1}^{2}\left(\frac{\partial
{\bf{r}}_i(z)}{\partial z}\right)^{^{\scriptstyle 2}} +
v_0\int_{0}^{N} dz\ V{\bf (}{\bf r}_{12}(z){\bf )} -
\frac{v_0^2\Delta}{2}\int_{0}^{N} dz
\ V^2{\bf (}{\bf r}_{12}(z){\bf )}. \label{eq:h1}
\end{equation}
It appears from the above expression of the effective
hamiltonian that an attraction is generated between the two
chains.  We find it instructive to follow another approach
of perturbation expansion of the interaction term starting
with the original hamiltonian ~\ref{eq:h}.  This helps us
in visualizing the origin of the disorder induced
attraction.  This is done in Appendix A.

Now, since any short range potential under renormalization
maps onto a $\delta$ function potential, we can take the
``minimal" effective hamiltonian for $\langle Z\rangle$ as
\begin{equation}
{\cal H}_{2}=\frac{1}{2} \int_{0}^{N} dz \
\left[\left(\frac{\partial
{\bf{r}}_1(z)}{\partial z}\right)^{^{\scriptstyle 2}}+
\left(\frac{\partial{\bf{r}}_2(z)}{\partial
z}\right)^{^{\scriptstyle 2}}\right]+
{\bar{v}}_0\int_{0}^{N} dz\ \delta{\bf (}{\bf
r}_{12}(z){\bf )}.
\label{eq:hm2}
\end{equation}
where ${\bar{v}}_0$ is the reduced coupling constant which
takes care of the attraction described earlier. We beleive
that the large length scale properties as described by
Eq.~\ref{eq:hm2} is same as that of Eq.~\ref{eq:h1}.  If
necessary, we can restrict the strength of the disorder so
that $\bar{v}_0$, which represents the effective coupling
between the two chains, is positive (i.e.  repulsive
interaction).  Now the problem reduces to a relatively
simple situation where the two chains interact with a pure
$\delta$-function interaction with a reduced coupling
constant ${\bar{v}}_0$.  The solution of this pure problem
is known, and is used below.\cite{jjb,smbjj}

For the sake of completeness we quote the relevant results
from from Ref.\cite{jjb}.  The perturbation series for the
connected part of the annealed partition function $\langle
Z\rangle_c$ to all orders in ${\bar{v}}_0$ is
\begin{equation}
\langle Z\rangle_c=N {\cal V} {\bar{v}}_0\ \left [ \ 1\
+\ \sum _{n=1}^{\infty}\ (-1)^n\
\frac{{\bar{v}}_0^n} {{(4 \pi)^{nd/2} }}
N^{n{\epsilon}'} \
\frac{\Gamma^n({\epsilon}')} {\Gamma
(2+n{\epsilon}')}\ \right ],\label{eq:zca}
\end{equation}
where ${\cal V}$ is the $d$-dimensional transverse volume, and
$\Gamma (\bullet)$ is the standard gamma function.
The exact $\beta$-function for the renormalized coupling
constant $u$ (with $u_0={\bar{v}}_0L^{2-d}$ as the bare
dimensionless coupling constant)
\begin{equation}
\beta(u)\equiv L\frac{\partial u}{\partial L}= 2 \epsilon' u
\left ( 1-\frac{u}{4\pi \epsilon'}\right )\label{eq:bu}.
\end{equation}
Note, here $2{\epsilon}'=(2-d)$ replaces $\epsilon$ of Ref
\cite{jjb} to
avoid later conflict of notation.

The flow diagram for the dimensionless coupling constant
$u$ is shown in Fig.~\ref{uflo}.  The fact that for $d<2$
any small attractive interaction is able to form a bound
state is reflected by the flow to the nonperturbative
regime for any negative $u$. The repulsive or the positive
$u$ region is dominated by the stable fixed point
$u^*(=4\pi{\epsilon}')$. For $d>2$ there exists a
nontrivial unstable fixed point $u=u^*$ which seperates the
bound and the unbound states for the two polymers. In
short, the unstable fixed point represents the critical
point for the binding-unbinding transition. The exponents
are known and can be found in Ref.
\cite{lip}. For example for $1\leq d< 4$, the length scale
exponent is $1/{\mid{\epsilon}'\mid}$.\cite{lip}

\section{$\langle Z^2\rangle$}
\label{sec:z2av}
The evaluation of $\langle Z^2\rangle$ closely parallels
that of the previous section.  However, unlike the $\langle
Z\rangle$ case, new terms are generated here in the
effective hamiltonian.  An RG analysis is done to get a
detailed account of the effects of these new terms.
\subsection{Effective hamiltonian}
\label{subsec:efham}
The averaging for $\langle Z^2\rangle$ with the hamiltonian
in Eq. ~\ref{eq:h} needs completion of the perfect square
associated with $b(z)$. As in the replica analysis
\cite{mpv} where
one needs $n$ replicas (``copies") of the original system
in evaluating $\langle Z^n\rangle$, we require four chains
for $\langle Z^2\rangle$, a pair \{3,4\} as a replica of
the original pair of chains \{1,2\}. Therefore, we write,
restricting ourselves to $\delta$-function potentials,
\begin{mathletters}
\begin{equation}
\langle Z^2\rangle =\int \prod_{i=1}^{4} Dr_i\ \ \exp (-{\cal
H}_{2,2}), \label{eq:zsa}
\end{equation}
where
\begin{equation}
{\cal H}_{2,2}=H_0+H_1+H_2,\label{eq:zsb}
\end{equation}
with
\begin{equation}
H_0=\frac{1}{2}\int_{0}^{N}dz\sum_{i=1}^{4}
\left (\frac{\partial{\bf{r}}_{i}(z)}{\partial z}\right
)^{^{\scriptstyle 2}}\label{eq:zsc}
\end{equation}
denoting the four chain free part, and
\begin{equation}
H_1=\bar{v}_0\int_{0}^{N}dz \ [\delta{\bf (}{\bf
r}_{12}(z){\bf )}+\delta{\bf (}{\bf r}_{34}(z){\bf
)}]\label{eq:zsd}
\end{equation}
and
\begin{equation}
H_2=-\bar{r}_0\int_0^N dz\ \delta{\bf (}{\bf r}_{12}(z){\bf
)}\
\delta{\bf (}{\bf r}_{34}(z){\bf )}\label{eq:zse}
\end{equation}
\end{mathletters}
representing the interactions among the chains, with
${\bar{r}}_0=v_0^2\Delta$.

In Eq. ~\ref{eq:zsb}, $H_1$ denotes the repulsive
interaction between the chains of a particular pair
(``intra replica", \{12\} and \{34\}, no cross coupling) at
the same chain length, and is identical to the interaction
term used for $\langle Z\rangle$ as discussed in section
{}~\ref{sec:zav} and Appendix A.  The other term $H_2$
couples the two pairs of chains $\{12\}$ and $\{34\}$
(``inter replica" term), and is the crucial term for our
analysis.  Eventhough this is a four chain interaction, it
is distinct from the multicritical type interaction of
Eq.~\ref{eq:hm}.  It cannot be interpreted directly as a
conventional interaction between the two pairs.  Rather,
there is a lowering of ``energy" of the system if the
partners of each pair \{12\} and \{34\} meet simultaneously
at the same chain length but not necessarily at the same
point in space.  This can also be interpreted as a special
correlation so that an encounter of \{12\} at a chain
length $z$ favors an encounter for \{34\} right at the same
length $z$. A tendency to achieve this kind of
configurations leads to all the nontrivial effects of the
disorder.

The coupling constant of $H_2$, ${\bar{r}}_0$ in Eq.
{}~\ref{eq:zse}, appears to be similar to that of the
attractive interaction which is present in $H_1$, Eqs.
{}~\ref{eq:zsd} and ~\ref{eq:h1}, but they require separate
treatments in the RG analysis (see below).  As discussed in
Appendix A, the term proportional to $v_0^2\Delta$ in
$\bar{v}_0$ is reduced by a cutoff volume factor $\Omega$
needed to define the $\delta^2$ term properly. Because of
this reduction ${v_0^2\Delta}/\Omega$ differs from
${\bar{r}}_0$ in dimensionality and matches properly with
$v_0$, the coupling constant of the starting $\delta$
function interaction of the two chains.

The standard dimensional analysis for dimensionless
hamiltonian shows that $[v_0]=L^{d-2}, [\Delta]=L^2$ and
hence $[v_0^2\Delta]=L^{2d-2}$ where $L$ has the dimension
of length.  Therefore, the upper critical dimension for
$\bar{r}_0$ is $d=1$ which also appears as a special
dimension through the divergences in the
$\epsilon(=1-d)$-expansion to be discussed below.  From
this simple dimensional analysis it also follows that the
coupling in $H_2$ differs from that of $H_1$, as already
mentioned.  The special dimensionality $d=1$ which is
associated with $H_2$ is more important in the context of
fluctuations in the partition function.  In the absence of
this term there is no special effect of disorder, which, in
turn, also means that the quenched and annealed free
energies would become equal.

\subsection{PERTURBATION SERIES}
\label{subsec:pert}

To study the effect of $H_2$, we develop a perturbation
series for $\langle Z^2\rangle$ in ${\bar{r}}_0$.  The
divergences that appear are absorbed by renormalization
through an $\epsilon$-expansion.  We show that this
renormalization can be carried out exactly to all orders.
For simplicity, this is done first for the $\bar{v}_0 = 0$
case.  We then show that these divergences at $d=1$,
arising only due to $H_2$, remain unchanged even if we
include $H_1$ i.e when ${\bar{v}}_0
\neq 0$.  The $\beta$ function
evaluated exactly to all orders in perturbation series and
other essential features are identical for both
${\bar{v}}_0 = 0$ and ${\bar{v}}_0
\neq 0$ cases.

\subsubsection{$\bar{v}_0=0$}
\label{subsubsec:zero}

Let us consider first, for simplicity, the case when
${\bar{v}}_0=0$.  This means that there is no mutual two
chain interaction.  We consider only the connected part
$\langle Z^2\rangle_c=\langle Z^2\rangle-\langle
Z\rangle^2$, the second cumulant of the partition function.
As in the previous section, the calculation can be done in
the real (chain) space.  But, at this point, we prefer the
Laplace space (Laplace transform with respect to the chain
length) because it is advantageous for later considerations
especially with $\bar{v}_0 \neq 0$. We define
\begin{equation}
{\cal Z} = \int_0^{\infty} dN \ e^{-sN} \langle
Z^2\rangle_c\label{eq:lap}
\end{equation}
the Laplace transform of $\langle Z^2\rangle_c$ with
respect to the chainlength $N$, the Laplace conjugate
variable being $s$.

The loops in the perturbation expansion are shown in Fig
{}~\ref{z2}a upto third order in the interaction.  The
individual pairs of chains are represented by thick lines.
The horizontal wiggly lines in these diagrams stand for
${\bar{r}}_0$. Such a representation is possible because
the $\delta$ function in $H_2$, Eq. ~\ref{eq:zse}, forces
the members of a pair to have the same ${\bf r}, z$
coordinates.  Each chain is described by the free
distribution (``propagator") $G({\bf r}_f - {\bf r}_i\mid
z_f - z_i) = [2\pi (z_f - z_i)]^{-d/2}
\exp [ - ({\bf r}_f - {\bf r}_i)^2/2(z_f - z_i) ]$ with end
points $({\bf r}_f,z_f)$ and $({\bf r}_i, z_i)$.  Two
chains are therefore described by
\begin{equation}
G^2 ({\bf r}\mid z) =(4\pi z)^{-d/2}\ G ({\bf r}\mid z/2)
.\label{eq:gsq}
\end{equation}
This $G^2$ is the propagator for the thick lines.  At each
wiggly line, connecting four chains (all four having the
same chainlength $z$), there are two integrations over the
spatial coordinates of the two separate pairs of chains
(thick lines).  The loops formed out of the wiggly lines
are only responsible for the divergences at $d=1$.

In order to trace the algebraic origin of the singularity,
note that, by very nature of the interaction, the spatial
integrations associated with the two thick lines are
independent of each other.  Each section of the thick
lines, with $z_1, z_2$ as the end points, in a loop formed
with the wiggly lines, contributes $(z_1 - z_2)^{-d/2}$
from the identity in Eq. ~\ref{eq:gsq}.  Since the
interaction demands same $z$ for the two thick lines, the
$z$ integrals involve $(z_1 - z_2)^{-d}$ type factors whose
Laplace transform would contribute $\Gamma(1-d)$ with pole
at $d=1$.  The two independent spatial coordinates which
are left out after the successive use of the normalization
$\int d{\bf r} \ G({\bf r}\mid z) = 1$, lead to a ${\cal
V}^2$ factor for each diagram. The convolution nature of
the $z$ integrals, thanks to the time ordering, leads to a
simple product of the individual Laplace transforms of the
integrands, resulting in a geometric series for ${\cal Z}$.
The details of the evaluation of a few diagrams are given
in Appendix A.

Defining the dimensionless coupling constant $r_0$ through
an arbitrary length scale $L$ as
$r_0=\bar{r}_0L^{2\epsilon}(4\pi)^{-d}$, $\epsilon=1-d$, we
write the series for ${\cal Z}$ to all orders in $r_0$ as
\begin{equation}
{\cal Z}\mid_{\bar{v}_0=0}=(4\pi)^d\ {\cal{V}}^2\ s^{-2} \
L^{-2\epsilon}\ \big [ r_0+
\sum_{n=1}^{\infty}r_0^{n+1}\ (sL^2)^{-n{\epsilon}}\
{\Gamma}^n(\epsilon)
\big].\label{eq:zf}
\end{equation}
It is clear from the above expression that there is a
divergence at $d=1$ at each order ($>1$). This is tackled
by renormalization below.

\subsubsection{$\bar{v}_0\neq 0$}
\label{subsubsec:nonz}
In the above analysis, we have taken ${\bar{v}}_0=0$.  We
now include ${\bar{v}}_0$ and show that the singularity
structure around $d=1$, as in Eq. ~\ref{eq:zf}, remains
unaffected.

In this case there are both intra-pair and inter-pair
interactions.  The intra pair interactions, i.e., the
mutual short range $\delta$-function interactions among the
members of the pairs are represented by dots on the thick
lines. The basic idea of the procedure adopted is to show
that the dots can be absorbed by dressing the
``propagators".  The original propagators (thick lines of
${\bar{v}}_0=0$ case) $G^2$ is modified by the dots but not
trivially.  The ``dressing" factor depends on whether the
chains are open or tied at the ends.  One therefore needs
two types of dressed propagators, $\bar{G}_M$ for the thick
lines in the loops, and $G_O$ for the same in the outer
legs not involved in the loops. (See Fig ~\ref{z2} c and
d.) Since these involve only two chain interactions, the
singularities are at $d=2$ of the type $\Gamma (1-d/2)$ as
in Eq.  ~\ref{eq:zca}.  These dressed propagators are to be
used, as appropriate, in the skeleton diagrams of the
${\bar{v}}_0=0$ case without the dots (see Fig ~\ref{z2}
a).  We just qoute the forms of these propagators below -
the details can be found in Appendix B.

For the one in which the two participating members of a
thick line are tied together at both the ends [$({\bf
o},0) and ({\bf r}, z)$] of the line, the form of the
dressed propagator with $n$ meetings (dots) is [Fig.
{}~\ref{z2}d ]
\begin{equation}
{\bar{G}}_{M}^{(n)} ({\bf r} \mid z) = (-{\bar{v}}_0)^n (4
\pi)^{-(n+1)d/2} \frac{\Gamma^{n+1}({{\epsilon}'})}
{\Gamma{\bf (}(n+1){\epsilon}'{\bf )}} z^{(n+1) {\epsilon}'
-1} G({\bf r} \mid z/2).\label{eq:gm}
\end{equation}
There is translational invariance in both ${\bf r}$ and
$z$.  With $\epsilon' = (2-d)/2$, this form surely reduces
to Eq.  ~\ref{eq:gsq} for $n=0$.  Similarly, for the other
type in which the two members of a thick line are tied
together only at one of the ends, integrations for the open
end coordinates need to be done.  The resulting dressed
propagator for chains of length $z$ with $n$ intermediate
dots (Fig ~\ref{z2}c) has the following form
\begin{equation}
{{G}}_{O}^{(n)}(z) =\frac{[-z^{{\epsilon}'} {\bar{v}}_0
\Gamma({\epsilon}')]^n}{\Gamma(1+n{\epsilon}')
(4\pi)^{nd/2}}.\label{eq:go}
\end{equation}
With $n=0$, $G_O^{(n)}(z) = 1$, as it should be, by the
normalization of the distribution function $G({\bf r}\mid
z)$.  Also $G_O^{(n)}(z)$ has no space dependence.

In a diagram of a particular order in ${\bar{r}}_0$, the
thick lines can have arbitrary order $n$ in ${\bar{v}}_0$
(i.e. arbitrary number of dots). All such diagrams
differing only in orders of ${\bar{v}}_0$ are combined
together by summing over $n$.  The full dressed propagators
are
\begin{equation}
{\bar{G}}_{M} ({\bf r} \mid z)= \sum_{n}{\bar{G}}_{M}^{(n)}
({\bf r} \mid z), \ \ {\rm and}\ \ {{G}}_{O}(z) = \sum_{n}
{{G}}_{O}^{(n)}(z).\label{gmgo}
\end{equation}

Unlike the ${\bar{v}}_0=0$ situation, these two propagators
replace the inner and the outer thick lines respectively.
The subsequent procedure is almost similar to the previous
case, including the origin of the ${\cal {V}}^2$ factor,
and the use of the convolution theorem in the Laplace
space.  The series for ${\cal Z}$ is given by
\begin{equation}
{\cal Z}\mid_{{\bar{v}}_0\neq 0} = {\cal V}^2 {\cal G}_O(s)
{\bar{r}}_0 \left [ 1 + \sum_{n=1}^{\infty}
\left ( \bar{r}_0 {\cal G}_M(s)\right )^n \right
]{\cal G}_O(s)\label{eq:lzg}
\end{equation}
where ${\cal G}_O(s)= {\cal L} G_O^2(z)$, ${\cal G}_M(s)=
{\cal L} G_M^2(z)$, (${\cal L}$ being the Laplace transform
with resprect to $z$) and $G_M(z) = \int d{\bf r}
{\bar{G}}_{M} ({\bf r} \mid z)$. The ``same $z$"
requirement of the $\delta$ function of $H_2$ combines the
propagators of the thick lines.  Hence $G_O^2$ and $G_M^2$.
The two outer pairs of legs contribute the two ${\cal
G}_0(s)$ factors.  In terms of the dimensionless coupling
constant $r_0$, and ${\bar{u}}_0 (= {\bar{v}}_0
(4\pi)^{-d/2}\Gamma(2{\epsilon}')s^{-{\epsilon}' })$, the
above expression becomes
\begin{equation}
{\cal Z}\mid_{{\bar{v}}_0\neq 0} =(4\pi)^d {\cal V}^2
s^{-2}L^{-2\epsilon} {\cal S}_O(s) \left [ r_0 +
\sum_{n=1}^{\infty} r_0^{n+1} (s L^2)^{-n\epsilon}
({\cal S}_M(s))^n \right ]{\cal S}_O(s)\label{eq:lzs}
\end{equation}
where
\begin{equation}
{\cal S}_0(s)= \sum_{\{n\}} \frac{
\Gamma[(n_1+n_2){\epsilon}' +
1]} {\Gamma (1+n_1 {\epsilon}') \Gamma (1+n_2 {\epsilon}')}
(-{\bar{u}}_0)^{n_1+n_2},\label{eq:zd}
\end{equation}
and
\begin{equation}
{\cal S}_M(s)=
\sum_{\{n\}}
\frac{\Gamma[(n_1+n_2){\epsilon}' + 1-d]}
{\Gamma {\bf (}(n_1+1) {\epsilon}'{\bf )}\Gamma{\bf
(}(n_2+1) {\epsilon}'{\bf )}}
(-{\bar{u}}_0)^{n_1+n_2}.\label{eq:mid}
\end{equation}
Details can be found in Appendix B.  The reason it is
written in the above form is that ${\cal S}_0$ and ${\cal
S}_M$ start with 1 for $n_1=n_2=0$, to agree with Eq.
{}~\ref{eq:zf}.  It follows that the leading divergences at
$d=1$ in each order of Eq.~\ref{eq:lzs} come from the $n_1
= n_2 =...=0$ term of Eq.~\ref{eq:zd} and Eq.~\ref{eq:mid}

\subsection{Renormalization and New Criticality}
\label{subsec:norm}
The divergences at $\epsilon=0$ in the series of
Eq.~\ref{eq:lzs} can be absorbed by the standard
renormalization procedure. \cite{amit} In general, a
renormalization through minimal subtraction would require
absorption of the poles in $\epsilon$ through
\begin{equation}
r_0 = r(1 + a_1 r + a_2 r^2 + ...).\label{eq:series}
\end{equation}
with $a_n = \sum_{p=1}^{n} a_{n,p} \epsilon^{-p}$ and $r$
as the renormalized coupling constant. In such a scheme,
$a_{n,p} (p \neq n)$ terms are required to take care of the
subleading divergences.

The formal similarity of the leading pole structure of Eqs
{}~\ref{eq:zf} and ~\ref{eq:lzs} with that of
Eq.~\ref{eq:zca} enables us to follow Ref.\cite{jjb,smbjj}
yielding $a_p = (-\epsilon)^{-p}$. The geometric series of
Eqs.  ~\ref{eq:lzs} and ~\ref{eq:zf} guarantees that the
removal of the leading poles is sufficient to remove the
subleading ones.  The presence of the dots (Fig ~\ref{z2}b)
through $n_1, n_2\neq 0$ in Eq.~\ref{eq:lzs} is felt
through the changes in the subleading divergences.  This
does not pose a problem and can, indeed, be checked
explicitly.  Note that ${\cal{S}}_M$ of Eq.~\ref{eq:mid}
has an expansion of the form
\begin{equation}
{\cal{S}}_M= \frac{1}{\epsilon} + A_0 + \sum_{p=1} A_p
\epsilon^p.
\end{equation}
Taking $a_p = (-\epsilon)^{-p}$, as needed to remove the
leading poles, one can verify explicitly that all the poles
are removed order by order, and the result does not depend
on the explicit values of $A_0, A_1$ etc.

The $\beta$ function is therefore exact to all orders in
perturbation series and is given by
\begin{equation}
\beta(r)\equiv L\frac{\partial r}{\partial L}=2 (\epsilon
r+r^2)\label{eq:br}.
\end{equation}
There are two fixed points: (i) $r=0$ and (ii)
$r^*=-\epsilon$. The flows are shown in Fig ~\ref{rflo}.
The bare coupling constant $r_0$ which originates from
$v_0^2\Delta$, where $\Delta$, the variance of the
distribution, is strictly positive, requires a positive
$r$. Therefore, the nontrivial fixed point for $d<1$ in
negative $r$ is unphysical.  It however moves to the
physical domain for $d>1$. See Fig. 3c.

Exactly at $d=1, \epsilon=0$, $r$ grows with length $L$ as
\begin{equation}
r(L)=r(0) \ \left [1+2 r(0)\ln\frac{L_0}{L}\right
]^{^{-1}},\label{eq:rl}
\end{equation}
$r(0)$ being the coupling at length $L_0$.  Hence, the
disorder is marginally relevant, in agreement with Ref
\cite{derhv}.  For
$d>1$, there exists an unstable nontrivial fixed point at
$r = \mid \epsilon \mid$ which separates two distinct
regimes of disorder. If we start with a strong enough
disorder, on the right side of the fixed point, it
increases with length scale, going beyond the perturbative
regime.  This is the strong disorder phase.  On the other
hand, the left side of the fixed point is the weak disorder
regime, since $r$ flows to zero (the stable fixed point).
The unstable fixed point, therefore, represents a critical
point - a novel phase transition induced by the disorder.

One way of achieving the abovementioned critical behavior
is to change the strength of the disorder by controlling
the temperature.  The ``strong disorder" phase $(\langle\ln
Z\rangle \neq
\ln \langle Z\rangle)$ would correspond to the low
temperature phase while
the ``weak disorder" phase $(\langle\ln Z\rangle\approx \ln
\langle Z\rangle)$ is the high
temperature one.  The details of the critical behavior can
be obtained by integrating the $\beta$ function,
\begin{equation}
r = \mid\epsilon\mid \left [1- \frac{r(0) - \mid
\epsilon\mid}{r(0)}
\left (
\frac{L}{L_0}\right )
{\raisebox{1.2em}{${2\mid \epsilon\mid}$}}
\right ] ^{^{-1}}.  \label{eq:rl2}
\end{equation}
For a small starting deviation $\Delta T \equiv T - T_c =
r(0) -
\mid \epsilon\mid$, there is a lengthscale $L \sim
(\Delta T)^{-1/2\mid\epsilon\mid}$ at which $r$ in Eq.
{}~\ref{eq:rl2} diverges.  This we can identify as a
lengthscale $\xi$ associated with the critical point with
the lengthscale exponent
\begin{equation}
\nu = (2\mid\epsilon\mid)^{-1}.\label{eq:nu}
\end{equation}
The divergence at $\epsilon=0$ is consistent with the
essential singularity that follows from Eq.~\ref{eq:rl}
\begin{equation}
\xi\sim\exp [1/(2\Delta T)]. \label{eq:xi}
\end{equation}

A complete description of the critical point would involve
an evaluation of various macroscopic or thermodynamic
properties. These would require a replica type analysis.
It is tempting to believe that the correlation induced by
$H_2$ in Eq.~\ref{eq:zse} in the replica space
distinguishes the two phases. We wish to come back to such
replica analysis elsewhere.

\section{$\langle Z^3\rangle$}
\label{sec:z3av}
The evaluation of $\langle Z^3\rangle$ leads to a six chain
problem where, as before, an interaction involving four
chains is generated which is attractive in nature. The
effective hamiltonian, apart from the free part for six
chains and mutual $\delta$ function interaction, contains
the following attractive terms (see Eq.~\ref{eq:zsb})
\[-{\bar{r}}_0\int_{0}^{N}dz\ [\delta{\bf (}{\bf
r}_{12}(z){\bf )}
\  \delta{\bf (}{\bf r}_{34}(z){\bf )}+
\delta{\bf (}{\bf r}_{34}(z){\bf )}\ \delta{\bf (}{\bf
r}_{56}(z){\bf )}+
\delta{\bf (}{\bf r}_{12}(z){\bf )}
\ \delta{\bf (}{\bf r}_{56}(z){\bf )}].\]

Instead of $\langle Z^3\rangle$, we analyze the third
cumulant $\langle Z^3\rangle_c$ involving only six chain
connected diagrams.  Fig ~\ref{z3} shows such diagrams upto
fourth order in ${\bar{r}}_0$.  Contributions of these
diagrams can be found out following the rules discussed in
the context of $\langle Z^2\rangle$ in section
{}~\ref{sec:z2av}.  As an example, we give an explicit
evaluation of Fig ~\ref{z3}c which is
\begin{eqnarray*}
{{\bar{r}}_0}^4\int_{\{{\bf r},{\bf r'},{\bf r''}\}}
\int_{0}^{N}dz_1\int_{0}^{z_1} dz_2\int_{0}^{z_2}
dz_3 \int_{0}^{z_3} dz_4 \ G^2({\bf r}_{12} \mid z_{12})\
G^2({\bf r'}_{12}
\mid z_{12})\times\\ G^2({\bf r'}_{23} \mid z_{23})
G^2({\bf r'}_{34} \mid z_{34})\ G^2({\bf r''}_{34}
\mid z_{34})
\end{eqnarray*}
In the above equation ${\bf r},\ {\bf r}',\ {\bf r}''$ with
appropriate subscripts denote the set of $d$-dimensional
coordinates for the three thick lines and $\int_{\{{\bf
r},{\bf r'},{\bf r''}\}}$ corresponds to the integrations
over all spatial coordinates. As before, each thick line
between two end points $({\bf r}_i , z_i)$ and $({\bf r}_j,
z_j)$ is represented by $ G^2({\bf r}_{ij}\mid z_{ij})$
with $z_{ij}=z_i-z_j$.  The spatial integrations simplifies
the above expression to
\begin{eqnarray*}
{\bar{r}_0}^4({4\pi})^{-5d/2}{\cal{V}}^3
\int_{0}^{N}dz_1\int_{0}^{z_1}
dz_2\int_{0}^{z_2} dz_3 \int_{0}^{z_3} dz_4 \ z_{12}^{-d}\
z_{23}^{-d/2}\ z_{34}^{-d}\nonumber
\end{eqnarray*}
\begin{equation}
= {\bar{r}_0}^4({4\pi})^{-5d/2}{\cal{V}}^3\frac{
{\Gamma}^2(1-d)\ \Gamma(1-d/2)
\ N^{4-5d/2}} {\Gamma(5-5d/2)}.
\end{equation}

The series for the Laplace transform of $\langle
Z^3\rangle_c$ is given by
\begin{eqnarray*}
\big[ {\cal{L}} \langle Z^3\rangle_c\big]
\mid_{{\bar{v}}_0=0}={\cal
V}^3({4\pi})^{3d/2}s^{-1-3d/2}\Gamma({\epsilon}')\big[
2{{r}_0}^ 2 (s L^2)^{-2\epsilon} + 4{{r}_0}^3(s
L^2)^{-3\epsilon}
\Gamma(\epsilon) +\nonumber
\end{eqnarray*}
\begin{equation}
6{{r}_0}^4(s
L^2)^{-4\epsilon}{\Gamma}^2(\epsilon)+...\big]\label{eq:z3a}
\end{equation}
where, as before, $s$ is the Laplace conjugate to the
chainlength $N$ and $r_0$ is the dimensionless coupling
constant as defined before Eq.~\ref{eq:zf}. This series
requires the standard renormalization procedure for removal
of divergence at $d=1$. Defining the renormalized $r$ via
\begin{mathletters}
\begin{equation}
r_0=r(1+a_1 r +a_2 r^2+...)
\end{equation}
it is found that
\begin{equation}
a_p=\big(-1/{\epsilon}\big)^p
\end{equation}
\end{mathletters}
absorbs the divergence at $d=1$.

It is interesting to note that at a particular order there
are diagrams which are similar by a mere permutaiton of the
interaction lines, i.e., by a different time ordering.  For
example, Fig ~\ref{z3}b shows four diagrams related by
permutations, in the third order of the perturbation
series.  All of these have the same value, and hence the
factor of 4 in the $r_0^3$ term in Eq. ~\ref{eq:z3a}.
These permutation factors collaborate with powers of $r_0$
in such a way that $a_p$'s are just the same as those for
the $\langle Z^2\rangle$ case.  There are also diagrams in
the third and higher orders (a few shown in Fig ~\ref{z3}d)
which correspond to subleading divergences, the removal of
which will be automatic by their corresponding higher
orders.

We, therefore, see that the $\beta$ function for $r$ has
exactly the identical form as that in Eq. ~\ref{eq:br} for
$\langle Z^2\rangle_c$ and all the features follow
identically. This shows that the phase transition for
$\langle Z^3\rangle_c$ has the same nature as for the
$\langle Z^2\rangle$ case.  To be more explicit, there
exists a transition temperature for $d>1$ which separates
the weak disorder and strong disorder phase for every
moment.  In the high temperature phase $\langle Z^3\rangle
\sim \langle Z\rangle^3$ and for
$T<T_c$, i.e., in the fluctuation dominated phase, $\langle
Z^3\rangle$ differs from $\langle Z\rangle^3$. This
transition temperature is the same for $\langle Z^3\rangle$
and $\langle Z^2\rangle$.

It is now a trivial exercise to extend this for higher
moments.\cite{rem}
The effective hamiltonian apart from the free part for $2 n$
chains and mutual $\delta$ function interaction involves the
following attractive interaction
\[-{\bar{r}}_0\sum_{i<j}\int_{0}^{N}dz\ \delta{\bf (}{\bf
r}_{2i-1\ 2i}(z){\bf )}
\  \delta{\bf (}{\bf r}_{2j-1\ 2j}(z){\bf )}\]
Since no new interaction is generated, the $\beta$ function
remains the same.\cite{path,amit}

\section{RANDOM MULTICRITICAL CASE}
\label{sec:multi}
In the previous sections, attention was focussed on the two
body interaction case. It is known that DPs with pure
$m$-body interaction can also be completely solved.
\cite{jjth,smbph,smbjj2} We now
investigate the random version of this multicritical case
as given by the hamiltonian of Eq.~\ref{eq:hm}.  As before,
we want to evaluate $\langle {Z}_m\rangle$ and $\langle
Z_m^2\rangle$.  The procedure follows the footsteps of the
two chain problem, and, therefore, details are skipped.

\subsection{$\langle Z_m\rangle$}
\label{subsec:zma}
To compute $\langle Z_m\rangle$, we can perform an
averaging over $b(z)$ to obtain, as in section
{}~\ref{sec:zav}, an $m$-chain hamiltonian with a pure
$m$-body interaction. The grand universality known for the
pure system indicates that the multicritical exponents for
the binding - unbinding transition will be similar to those
of Ref \cite{lip}. For example, for $d>2/(m-1)$, the
lengthscale exponent would be $2/{\mid{\epsilon}_m '\mid}$,
where ${\epsilon}_m'=2-(m-1)d$.  '

\subsection{$\langle Z_m^2\rangle$}
\label{subsec:zm2}
A little calculation involving the completion of the square
would convince the reader that the effective hamiltonian
needed for the second moment would involve $2 m$ chains in
sets of $m$.  It is given by
\begin{displaymath}
{\cal H}_{m,m} = \frac{1}{2}\ \int\limits_{0}^{N} \ dz
\sum_{i=1}^{2m} \left (\frac{\partial
{\bf{r}}_i(z)}{\partial z}\right )^{^{\scriptstyle 2}} +
{\bar{v}_m}
\int\limits_{_{0}}^{N}\ dz \
\prod_{p=1}^{m-1}\ \delta [{\bf r}_{p\ p+1} (z)]\ +
\end{displaymath}
\begin{equation}
{\bar{v}_m} \int\limits_{_{0}}^{N}\ dz \
\prod_{q=m+1}^{2m-1}\ \delta [{\bf r}_{q\ q+1} (z)]
\ -
{{\bar{r}}_m} \int\limits_{_{0}}^{N}\ dz \
\prod_{p=1}^{m-1}\ \delta [{\bf r}_{p\ p+1} (z)]\
\prod_{q=m+1}^{2m-1}\ \delta [{\bf r}_{q\ q+1} (z)]
\label{eq:h2m}
\end{equation}
where ${\bar{r}}_m=v_m^2\Delta$ and the two sets are
represented by $p$'s and $q$'s.  The special feature is the
last term that involves the peculiar $m$ chain - $m$ chain
interaction. This generalizes $H_2$ of the two chain case
of Eq. ~\ref{eq:zsb} and ~\ref{eq:zse}. The effect of
disorder, so far as the fluctuations are concerned, is to
introduce a correlation that if $m$ chains meet at a $z$,
the replica would also like to enjoy a meeting at that same
$z$.

The upper critical dimension of $r_m$ follows from the
dimensional analysis as $d_m=1/(m-1)$, which is half of the
upper critical dimension for the pure case [$2/(m-1)$ for
${\bar{v}}_m$]. We are not sure whether this systematic
reduction by a factor of $2$ has any deeper significance.

For simplicity we choose $\bar{v}_m=0$. The perturbation
expansion in ${\bar{r}}_m$ would involve the same sets of
diagrams as in Fig. ~\ref{z2}a except that the propagator
now for the thick lines is $G^m({\bf r}\mid z)=(2\pi
z)^{-(m-1)d/2}m^{-d/2} G({\bf r}\mid z/m)$.  With this
propagator, the full series can be computed. It is
transparent to see the occurence of divergences at $d=d_m$.
The whole RG procedure of section ~\ref{sec:z2av} can be
copied in toto by replacing $\epsilon$ by ${\epsilon}_m=
1-d(m-1)$. Hence in the multicritical situation we also
expect to see a disorder induced phase transition. The
length scale exponent $\nu_m= (2\mid\epsilon_m\mid)^{-1}$,
with an essential singularity for $d=d_m$ as in
Eq.~\ref{eq:xi}.

\section{System with more than two chains}
\label{sec:big}
Annealed averaging for the system with two chains with
random interaction is simpler and not sufficient to give
enough information about the effects of disorder. On the
otherhand, if the above case can be extended to four chains
having twobody interaction among each other, even the
annealed case turns out to be extremely nontrivial.  The
hamiltonian for the four chain system,
\begin{equation}
{\sf H}=\frac{1}{2}\int_{0}^{N}dz\sum_{i=1}^{4}
\left (\frac{\partial{\bf{r}}_{i}(z)}{\partial z}\right
)^{^{\scriptstyle 2}} +
\int_{0}^{N} dz v_0(1+b(z))\sum_{\begin{array}{c} i,j=1
\\ i\neq j\end{array}}^4 \delta({\bf r}_{ij}(z))
\end{equation}
where ${\bf r}_{ij}={\bf r}_i(z)-{\bf r}_j(z)$, after
averaging, using the Gaussian distribution of $b(z)$, gives
the following effective hamiltonian
\begin{eqnarray}
{\cal H}_{{\rm
eff}}=&\frac{1}{2}\int_{0}^{N}dz\sum_{i=1}^{4}
\left (\frac{\partial{\bf{r}}_{i}(z)}{\partial z}\right
)^{^{\scriptstyle 2}} +
\left(\frac{\partial{\bf{r}}_{i}(z)}{\partial
z}\right)^{^{\scriptstyle 2}} +
\bar{v}_0\int_{0}^{N} dz \sum_{i,j=1}^4
\delta({\bf r}_{ij}(z))
-\nonumber \\ &2v_0^2\Delta
\int_{0}^{N}\sum_{i,j,k}\delta({\bf r}_{ij})
\delta({\bf r}_{jk})
- 2v_0^2\Delta\int_{0}^{N} dz
\sum_{i\neq j\neq k\neq l}
\delta({\bf r}_{ij}) \delta({\bf r}_{kl}).\label{eq:h4}
\end{eqnarray}
The remarkable feature of the effective hamiltonian is that
there are two different kinds of attractive interaction one
of which involves three chains with a multicritical type
interaction (Eq. ~\ref{eq:hm}) while the other one coupling
four chains together, as in the quenched problem.

If we take a three chain system, the corresponding
effective hamiltonian will involve only the three chain
term but no four chain interaction of Eq. ~\ref{eq:h4}.
This term was absent in the original three chain
hamiltonian.  There is now the possibility of a disorder
induced multicritical behavior, though of pure
type.\cite{jjth,smbph}

The four chain attractive interaction is marginal at $d=1$
and so is the three chain interaction.  The presence of
these two marginal operators is, in general, expected to
complicate the renormalization procedure through their
interdependence - but here that does not happen.

The perturbation expansion with the three body and the four
body interactions leads to three different kinds of
diagrams.  See Fig ~\ref{zann}.  The series corresponding
to the pure three body interaction is already
solved.\cite{jjth,smbph} The series in the Laplace space
involving four body interactions (see Fig 5a-c), which
contributes to the leading divergence is identical to the
series for $\langle Z^2\rangle_c$ in Eq. ~\ref{eq:zf} (Fig.
{}~\ref{z2}a). The diagrams with mixed three body and four
body interactions are shown in Fig ~\ref{zann}d,e.  In the
final series, upto the order shown in Fig ~\ref{zann}, the
four body and three body contributions get separated into
two factors.  This shows that the resulting renormalization
of the two couplings are independent of each other.
Because of the four body interaction, we expect a disorder
induced criticality as for the two chain quenched case, but
here this happens for a real four chain system - no replica
is involved.  The details and the phase diagram will be
published elsewhere.
\section{Summary and discussion}
\label{sec:smry}
We have proposed a random interaction model for two
directed polymers and studied the first three cumulants of
the partition function. We have shown that in the annealed
case, described by $\langle Z\rangle$, there can be a
disorder induced binding unbinding transition, very similar
to a pure problem. The exponents are also identical to the
pure case.  We also pointed out certain peculiarities of
the annealed problem involving three or four chains.  The
quenched problem is different as reflected through the
marginal relevance of the disorder. For $d>1$ there exists
a critical point that demarcates a disorder dominated phase
and a pure type phase. In the strong disorder phase, there
seems to have an extra correlation in the replica space,
which is absent in the other phase. The lengthscale
exponent for the critical point is found to be $(2\mid
\epsilon\mid)^{-1}$ where $\epsilon=1-d$.  It
has an exponential divergence at $d=1$. Similar results
were obtained for $\langle Z^3\rangle$. In the replica
approach, one needs $\langle Z^n\rangle$ with $n\rightarrow
0$ which does not require interactions other than those
which are present in $\langle Z^2\rangle$ and $\langle
Z^3\rangle$. Therefore the upper critical dimension will
remain the same, namely $d=1$.

There are still many open problems, as for example, a
replica analysis for this system.  This requires an
explicit expression for $\langle Z^n\rangle$, correct at
least for small $n$.  Such an analysis would provide vital
information regarding the disorder induced critical point,
including possible replica symmetry breaking.  Is there any
other length scale exponent for this new crtical point,
apart from the one we have calculated? What about other
exponents? What, if any, is the upper critical dimension of
this critical point?  A thorough numerical study of this
system will surely provide valuable insights.

\appendix{Generation of attraction: \\ Perturbation analysis
for $\langle Z\rangle$ and $\langle Z^2\rangle$}

In this appendix, we show how the attractive terms in the
effective hamiltonian for $\langle Z\rangle$ and $\langle
Z^2\rangle$ can be generated perturbatively.

We proceed to the evaluation of the average of the
partition function $\langle Z\rangle$ by a perturbation
expansion using Eq.  ~\ref{eq:h} with the replacement of
the short range potential by a $\delta$ function potential.
Formally this leads to the expression for
\begin {equation}
Z=\int Dr_1Dr_2 \exp{(-H_0)}\
\big(1-H_i+{{H_i}^2}/{2!}+...\big)\label{eq:zp}
\end{equation}
where $H_0$ corresponds to the free part of the two chains,
and $H_i$ represents the interaction part, $\int_0^N dz \
v_0\ (1+b(z)) \delta({\bf r}_{12}(z))$.  A little
manipulation after substituting the explicit form of $H_i$
gives the connected partition function as
\begin{eqnarray}
Z_c = & \int_{0}^{N} dz_1\ {v_0\ (1+b(z_1))}
\int_{\{r,r'\}}
G({\bf r}_N-{\bf r}_1\mid N-z_1)\ G({\bf r}_1-{\bf r}_0\mid
z_1)\times\nonumber\\ & G({\bf r'}_N-{\bf r}_1\mid N-z_1)\
G({\bf r}_1-{\bf r'}_0\mid z_1) +\nonumber\\ & \int_{0}^{N}
dz_1\
\int_{0}^{z_1} dz_2\
{v_0^2\ (1+b(z_1))} {(1+b(z_2))}\times\nonumber\\ &
\int_{\{r,r'\}} G({\bf r}_N-{\bf r}_1\mid N-z_1)\ G({\bf
r}_{12} \mid z_{12})
\ G({\bf r}_2-{\bf r}_0\mid z_2) \times\nonumber\\
& G({\bf r'}_N-{\bf r}_1\mid N-z_1) G({\bf r}_{12}\mid
z_{12})G({\bf r}_2-{\bf r'}_0\mid z_2)+...\label{eq:a2}
\end{eqnarray}
where $G({\bf r}\mid z)= {(2\pi z)}^{(-d/2)}
\exp (-r^2/2z)$ is the distribution function for a
$d$-dimensional Gaussian chain of chain length $z$ and
end-end distance vector ${\bf r}$ and $\int_{\{r,r'\}}$ the
integrations over all dummy spatial coordinates. The
convention is to use ${\bf r}$ and ${\bf r}'$ for the two
chains and $z_{ij}=z_i-z_j$.  The factorials in the
denominators of the terms of Eq. ~\ref{eq:zp} are absorbed
by introducing ``time" $(z)$ ordering, which restricts
$z_{i+1} \leq z_i$, in the integrals.  Diagrams upto second
order corresponding to the series in Eq.  ~\ref{eq:a2} are
shown in Fig ~\ref{zav}.  The familiar normalization $\int
d{\bf r} G(r\mid z)=1$ and the integrations over the
spatial end coordinates lead to much more simplification
for $Z_c$.  The explicit form of $Z_C$ upto two loop term
is given by
\begin{eqnarray}
& Z_C={\cal{V}}\int_{0}^{N} dz_1{v_0(1+b(z_1))} -
\int_{0}^{N} dz_1\int_{0}^{z_1}
dz_2 \int_{\{r\}} {v_0^2\ (1+b(z_1))}{(1+b(z_2))}
\nonumber \times\\
& G^2({\bf r}_{12}\mid z_{12})+
\int_{0}^{N} dz_1\int_{0}^{z_1}dz_2 \int_{0}^{z_2} dz_3
\int_{\{r\}}
v_0^3(1+b(z_1)) (1+b(z_2))\times \nonumber\\ & (1+b(z_3))
G^2({\bf r}_{12}\mid z_{12}) G^2({\bf r}_{23}\mid
z_{23})....\label{eq:zp2}
\end{eqnarray}
where ${\cal V}$ is the transverse volume.  The actual
meaningful quantity $\langle Z_c\rangle$ is computed from
Eq.  ~\ref{eq:zp2} after averaging it with the distribution
$P(b)$ of Eq. ~\ref{eq:disa}. This gaussian distribution
with zero mean ensures that any term involving odd number
of $b(z)$'s should vanish after averaging.  Therefore, the
contribution from the first order term, $v_0N {\cal{V}} $,
is only from the pure part.  In the one loop level of Eq.
{}~\ref{eq:zp2}, there are two surviving terms after disorder
averaging.  One is the pure term which does not require any
averaging and its contribution to $\langle Z_c\rangle$ is
\begin{eqnarray}
- v_0^2 {\cal V} (4\pi)^{-d/2}
\frac{\Gamma(1-d/2)}{\Gamma(3-d/2)}
N^{2-d/2}.\label{eq:t1}
\end{eqnarray}
The other nonvanishing part which contains an even number
of disorder interaction is
\begin{eqnarray}
- {v}_0^2\int_{0}^{N} dz_1\int_{0}^{z_1}\nonumber dz_2 \
\int
d{\bf r}_1\ d{\bf r}_2 \ b(z_1)\ b(z_2)\ G^2({\bf
r}_{12}\mid z_{12}).\nonumber
\end{eqnarray}
After averaging, the two points $z_1$ and $z_2$ along the
chain merge together giving rise to a term
\begin{eqnarray}
-\int_{0}^{N}dz_1\ v_0^2\ \Delta\delta^2({\bf r}_1-{\bf
r}_2)\ d{\bf r}_1 \ d{\bf r}_2\label{eq:t22}
\end{eqnarray}
where we used the fact that $G({\bf r}\mid 0) = \delta
({\bf r})$.  Because of this merging of the two points
along the chain, this term contributes to the first order
term but with a negative sign which shows the presence of a
newly generated attraction.  In other words, a second order
term for a particular realization (before averaging), looks
like an attractive first order term after DA.

Proceeding in the same fashion we can evaluate the two loop
term of Eq. ~\ref{eq:zp2}.  In the two loop part, the
nonvanishing contributions are:
\begin{eqnarray}
(i)\ \ v_0^3\int_{0}^{N} dz_1\int_{0}^{z_1}\nonumber
dz_2\int_{0}^{z_2}dz_3 {\int}_{\{r\}} G^2({\bf r}_{12}\mid
z_{12}) G^2({\bf r}_{23}\mid z_{23}).
\end{eqnarray}
This term involving only pure type interaction, after
integration over spatial coordinates, gives
\FL
\begin{eqnarray}
{\cal V} v_0^3 (4\pi)^{-d}\int_{0}^{N}
dz_1\int_{0}^{z_1}\nonumber dz_2\int_{0}^{z_2}dz_3\
(z_1-z_2)^{-d/2}\nonumber (z_2-z_3)^{-d/2}\\ ={\cal V}
v_0^3 (4\pi)^{-d}
\frac{\Gamma^2(1-d/2)}{\Gamma(4-d/2)}N^{3-d/2}.
\end{eqnarray}

\begin{eqnarray}
(ii)v_0^3\int_{0}^{N}dz_1\int_{0}^{z_1}\nonumber
dz_2\int_{0}^{z_2}dz_3 {\int}_{\{r\}}\ b(z_1) b(z_2)\
G^2({\bf r}_{12}\mid z_{12}) G^2({\bf r}_{23}\mid
z_{23}).\label{eq:t3}
\end{eqnarray}
As in the earlier case this reduces to a one loop term
after averaging as
\begin{eqnarray}
v_0^3\Delta\ (4\pi)^{-d/2}\int_{0}^{N}dz_1\int_{0}^{z_1}
dz_2\
\int_0^{z_2}dz_3 \int_{\{r\}}{\delta}^2({\bf r}_{12})\
(z_2-z_3)^{-d/2}.\label{eq:t23}
\end{eqnarray}
(iii) The third nonvanishing contribution is from
\begin{equation}
v_0^3\int_{0}^{N} dz_1\int_{0}^{z_1} dz_2\int_{0}^{z_2}dz_3
\int_{\{r\}}b(z_2)\ b(z_3)\ G^2({\bf r}_{12}
\mid z_{12})\ G^2({\bf r}_{23}\mid z_{23}).
\end{equation}
This term after averaging becomes
\begin{equation}
v_0^3 \Delta\ (4\pi)^{-d/2}\int_{0}^{N}dz_1\int_{0}^{z_1}
dz_2\
\int_{\{r\}} {\delta}^2({\bf r}_{23}) \
(z_1-z_2)^{-d/2}\label{eq:del3}.
\end{equation}
There is one more term involving $b(z_1)b(z_3)$ which
vanishes after averaging because of the specific time
ordering which rules out the merging of $z_1$ and $z_3$.
It can be easily be checked that this merging of two
``random interaction" lines (wavy lines) into one single
pure line and the subsequent reduction of order occurs at
each order ($>1$) involving consecutive pairs of even
number of ``random" lines.  Thus an attraction is generated
at each order very systematically.  The coupling constant
of this term is proportional to ${v_0}^2\Delta$.  In some
of the above expressions, [~\ref{eq:t22}, ~\ref{eq:t23},
{}~\ref{eq:del3}] the presence of the ${\delta}^2$ term needs
special attention since it is illdefined even in the theory
of distribution sense.
\cite{dist} We can avoid this problem by
taking a spread out $\delta$ function, and then taking the
limit at the end.  This would change the coupling to $v_0^2
\Delta/\Omega$, where $\Omega$ is the arbitrary ``spread
out" or cutoff volume.  Since one gets back a single
$\delta$ function, it can be associated with the pure term,
thereby changing the problem to a pure one with a reduced
coupling constant ${\bar{v}_0} = v_0 - v_0^2\Delta/\Omega$.
See Eq.  ~\ref{eq:h1}.  Another way to tackle this
difficulty is to start with a short range potential $V({\bf
r})$, and appeal to RG arguments as done in section
{}~\ref{sec:zav}.

{}From now onwards it's time for $\langle Z^2\rangle$ A few
diagrams for $\langle Z^2\rangle$ with ${\bar{v}}_0=0$ are
shown in fig. 7a.  In the first order, the only diagram
which has nonzero contribution to $\langle Z^2\rangle_c$ is
Fig. 7a1.  This contribution after disorder averaging is
${\cal V}^2 v_0^2 \Delta N$. As was mentioned in the text
this ${\cal V}^2$ has come from the independent spatial
integrals.  Proceeding in the similar fashion, we write
\begin{eqnarray}
fig 7a3 =v_0^4\int_{0}^{N} dz_1 \int_{0}^{z_1}dz_2\
b(z_1)b(z_2)\ \int_{\{r\}} G^2({\bf r}_{12}\mid
z_{12})\nonumber
\times\\
\int_{0}^{N}dz'_1\nonumber
\int_{0}^{z'_1}dz'_2 b(z'_1)b(z'_2)\int_{\{r, r'\}}
G^2({\bf r'}_{12}\mid z'_{12}).
\end{eqnarray}
In disorder averaging, the only relevant contribution comes
from the pairing of $b(z_1)$ $b(z_1')$ and $b(z_2)$
$b(z'_2)$.  The other possibility in which $b(z_1)$
$b(z_2)$ and $b(z'_1)$ $b(z'_2)$ are paired up (Fig 7b2) is
not considered here since this generates ${\bar{v}}_0$ type
terms which are not to be included for the ${\bar{v}}_0= 0$
case. After appropriate disorder averaging the above
expression becomes
\begin{equation}
v_0^4{\Delta}^2{\cal{V}}^2
{4\pi}^d\Gamma(\epsilon)s^{-(2+\epsilon)},\label{eq:a10}
\end{equation}
where $\epsilon=1-d$.

To make the evaluation, after DA, easier we follow a
different convention for the diagrams, Fig ~\ref{z2av}b and
2a.  The thick line represents the two members of a pair
jointly and is represented by $G^2({\bf r}_1 - {\bf r}_2
\mid z_1 -z_2)$
for the ends ${\bf r}_1$ and ${\bf r}_2$ at which the two
chains are tied at lengths $z_1$ and $z_2$.  For example,
the diagram which corresponds to the last expression (Eq.
{}~\ref{eq:a10}) is given by Fig ~\ref{z2av}b3, which is also
Fig 2a2. (Note ${\bar{r}}_0 = v_0^2\Delta$.)

The next diagram which is important in the next higher
order is given in Fig 7a4, the contribution of which can
equivalently be calculated from Fig 7b5 or 2a3 as
\begin{eqnarray}
(v_0^2\Delta)^3
\int_{0}^{N}dz_1\int_{0}^{z_1}dz_2\int_{0}^{z_2}
dz_3 \nonumber
\int_{\{r,\ r'\}}G^2({\bf r}_{12}\mid z_{12})
G^2({\bf r}_{23}\mid z_{23})\nonumber
\times\\
G^2({\bf r'}_{12}\mid z_{12}) G^2({\bf r'}_{23}\mid
z_{23}).
\end{eqnarray}
In the Laplace space this becomes
\begin{equation}
{(v_0^2\Delta)^3{\cal{V}}^2\Gamma^2(\epsilon)}
{(4\pi)^{-2d}s^{-(2+2\epsilon)}}
\end{equation}
The diagrams having odd numbers of wiggly lines trivially
vanishes after DA. This can be generalized to arbitrary
orders since only ladder type diagrams are involved.  Eq.
{}~\ref{eq:zf} would follow by substitution ${\bar{r}}_0=
v_0^2\Delta$.

\appendix{Dressed propagators and $\langle Z^2\rangle$}

We first show the two different dressed propagators.  The
one for which both the chains, tied at the ends $({{\bf
r},z})$ and $({{\bf r'},z'})$, meet each other $n$ times at
$({\bf r}_1, z_1)$, $({\bf r}_2, z_2)$, .....$({\bf r}_n,
z_n)$ (Fig ~\ref{z2}d) is given by
\begin{eqnarray*}
& {\bar{G}}_M^{(n)}({\bf r}-{\bf r}'\mid
z-z')={\bar{v}}_0^n\int_{\{\bf{r}\}} \int_{z'}^{z}
{dz_1}\int_{z'}^{z_1}dz_2
...\int_{z'}^{z_{n-1}}dz_n G^2({\bf r}-{\bf r}_1\mid z-z_1)
\times\\
& G^2({\bf r}_{12}\mid z_{12}).... G^2({\bf r}_{n-1n}\mid
z_{n-1n}) G^2({\bf r}_n-{\bf r}'\mid z_n-z')
\end{eqnarray*}
Use of the identity $ G^2({\bf r}\mid z)=(4 \pi
z)^{-d/2}G({\bf r}\mid z/2)$ and the Markovian property
\[\int d{\bf r}_2 G({\bf r}_1-{\bf r}_2 \mid z_1)
G({\bf r}_2-{\bf r}_3\mid z_2) =G({\bf r}_1-{\bf r}_3\mid
z_1+z_2)\] leads to the following expression for ${\bar
{G}}_M^{(n)}(r-r'\mid z-z')$.

\begin{eqnarray*}
& {\bar{G}}_M^{(n)}({\bf r}-{\bf r'}\mid
z-z')={\bar{v}}_0^n{4\pi}^{-nd/2} G({\bf r}-{\bf r}'\mid
(z-z')/2)\times\nonumber\\ & \int_{z'}^{z}
{dz_1}\int_{z'}^{z_1}dz_2
...\int_{z'}^{z_{n-1}}dz_n  (z-z_1)^{-d/2}
z_{12}^{-d/2}....\times\nonumber
\\&  z_{n-1n}^{-d/2} (z_n-z')^{-d/2}
\end{eqnarray*}
So we need the following integral
\begin{eqnarray*}
\int_{z'}^{z} {dz_1} (z-z_1)^{-d/2}\int_{z'}^{z_1}dz_2
(z_1-z_2)^{-d/2}...
\int_{z'}^{z_{n-1}}dz_n (z_{n-1}-z_n)^{-d/2}
(z_n-z')^{-d/2}.
\end{eqnarray*}
A change of variable
\[\bar{z}_i=z_i-z'\] and use of the convolution theorem in
the Laplace space straightaway yields
\[ s^{-{{\epsilon}'(n+1)}}{\Gamma}^{n+1}({\epsilon}')\]
where $s$ is the Laplace conjugate to the chainlength (more
precisely to ${\bar{z}}=z-z'$).  Converting this to inverse
Laplace space and combining all other factors, the final
form of such a propagator becomes
\begin{equation}
{\bar{G}}_{M}^{(n)}({\bf r}\mid z) =(-{{\bar{v}}_0})^n
(4\pi)^{-(n+1)d/2}z^{(n+1){\epsilon}'-1}
\frac{\Gamma^{n+1}({\epsilon}')}{\Gamma[(n+1){\epsilon}']}\
G({\bf r}\mid z/2).
\end{equation}
Following the same track, the other dressed propagator, for
which the two member chains are tied only at one end say
$({\bf r},\ z)$ other than $n$ meetings, has the form (see
Fig 2c)
\begin{equation}
G_{O}^{(n)}(z)=(-{{\bar{v}}_0})^n(4\pi)^{-nd/2}(z)^{n
{\epsilon}'}\frac{
\Gamma^n({\epsilon}')}
{\Gamma(n{\epsilon}'+1)}
\end{equation}
This propagator is independent of any space coordinate
because of the spatial integration over the open end
coordinates.

To take care of arbitrary number of meetings we sum over
$n$. Hence the final dressed propagators are
\[G_O(z)=\sum_{n=0}^{\infty}G_O^{(n)}(z)\]
\[{\bar {G}}_M({\bf r}-{\bf r}'\mid z-z')=\sum_{n=0}^
{\infty}{\bar{G}}_M^{(n)}({\bf r}-{\bf r}'\mid z-z')\]
Therefore the series for $\langle Z^2\rangle_c$ can be
written as
\begin{equation}
\langle Z^2\rangle_c={\bar{r}}_0\int_{0}^{N} dz_1
G_O^2(N-z_1) \ G_O^2(z_1) +\nonumber
\end{equation}
\begin{equation}
{\bar{r}}_0^2\int_{0}^{N} dz_1\int_{0}^{z_1}dz_2
{\int}_{\{r,\ r'\}} G_O^2(N-z_1)\
{\bar{G}}_M({\bf{r}}_{12}\mid z_{12})\
{\bar{G}}_M({\bf{r}}'_{12}\mid z_{12})\
G_O^2(z_2) + ... .
\end{equation}
In fact, it is possible to write the whole series to all
orders in perturbation. Completing the integrations over
the spatial coordinates we get
\begin{eqnarray}
& \langle Z^2\rangle_c={\cal V}^2{\bar{r}}_0\int_{0}^{N}
dz_1 G_O^2(N-z_1) G_O^2(z_1) + {\bar{r}}_0^2{\cal
V}^2\int_{0}^{N} dz_1\int_{0}^{z_1}dz_2 G_O^2(N-z_1)\times
\nonumber
\\
& G_M^2(z_{12}) G_O^2(z_2) +
{\bar{r}}_0^3{\cal{V}}^2\int_{0}^{N}
dz_1\int_{0}^{z_1}dz_2\int_{0}^{z_2} dz_3\
G_O^2(N-z_1)\times\nonumber
\\ & G_M^2(z_{12}) G_M^2(z_{23}) G_O^2(z_3).\label{eq:z2cr}
\end{eqnarray}
Here $G_M(z)$ is obtained from ${\bar{G}}_M({\bf r}\mid z)$
after integration over the spatial coordinate.  We use the
convolution theorem for Laplace transforms which leads to
the following expression for ${\cal Z}$ the Laplace
transform of $\langle Z^2\rangle_c$ (Eq. ~\ref{eq:lap})
\begin{eqnarray}
{\cal Z}\mid_{{\bar{v}}_0 \neq 0} = {\cal V}^2 [{\bar{r}}_0{\cal
G}_O^2(s) + {\bar{r}}_0^2 {\cal G}_O(s){\cal
G}_M(s)\nonumber {\cal G}_O(s)+\\ {\bar{r}}_0^3 {\cal
G}_O(s){\cal G}_M^2(s) {\cal G}_O(s)+....] .\label{eq:z2cl}
\end{eqnarray}
where, ${\cal Z} = \int_0^{\infty} e^{-sN} \langle
Z^2\rangle_c$, ${\cal G}_p = \int_0^{\infty} e^{-sN}
G_p^2(z)$, with $p$ being $O$ or $M$.  The Laplace
transforms ${\cal G}_O (s) $ and ${\cal G}_M (s)$ are given
by
\begin{equation}
{\cal G}_O(s)=\sum_{n_1,n_2}\frac{(4\pi)^{-(n_1+n_2)d/2}
{\bar{v}}_0^{\ n_1+n_2} \Gamma^{n_1+n_2} ({\epsilon}')
\Gamma((n_1+n_2){\epsilon}'+1)}{\Gamma(1+n_1{\epsilon}')
\Gamma(1+n_2{\epsilon}')\ s^{(n_1+n_2)(1-d/2)+1}}
\end{equation}
and
\begin{equation}
{\cal G}_M(s)=\sum_{n_1,n_2}\frac{(4\pi)^{-(n_1+n_2+2)d/2}
{\bar{v}}_0^{\ n_1+n_2}
\Gamma^{n_1+n_2}({\epsilon}')
\Gamma[(n_1+n_2){\epsilon}'+\epsilon]}
{\Gamma[(1+n_1){\epsilon}']
\Gamma[(1+n_2){\epsilon}']\ s^{(n_1+n_2){\epsilon}'+\epsilon}}.
\end{equation}
Substituting these expressions in equation ~\ref{eq:z2cl},
we get back equation ~\ref {eq:lzs} for ${\cal
Z}\mid_{{\bar{v}}_0\neq 0}$ in the Laplace space.  The
results for ${\bar{v}}_0 = 0$ can be obtained from the
first term of each sum, i.e., for $n_1 = n_2 =0$.

\figure{Flow diagrams for coupling constant $u$ in
different dimensions. $u^* (= 4\pi \epsilon')$ represents
the nontrivial fixed point.\label{uflo}}
\figure{(a) The only contributing diagrams in $\langle
Z^2\rangle\mid_{{\bar{v}}_0=0}$ upto third order.  Only
ladder diagrams occur. A thick line corresponds to a pair
of chains. A wiggly line stands for an ${\bar{r}}_0$ factor
in the evaluation of the diagrams. (b) A typical diagram
for $\langle Z^2\rangle\mid_{{\bar{v}}_0\neq0}$.  The dots
on the thick lines represent intrapair interactions
$({\bar{v}}_0)$. (c) The dressed propagator with two chains
tied at only one end.  The dashed lines represent the
mutual $\delta$-function type interaction with coupling
constant ${\bar{v}}_0$. (d) The dressed propagator with two
chains tied at both $({\bf r}, z)$, and $({\bf r}', z')$.
\label{z2}}
\figure{Flow diagram for $r$ in various dimensions.
$r^* (= -\epsilon)$ represents the nontrivial unstable
fixed point.  For the $m$th order multicritical case $r$ is
to be replaced by $r_m$.  The three figures would be for
$d<d_m$, $d=d_m$, and $d>d_m$, where $d_m = 1/(m-1)$.
\label{rflo}}
\figure{The second (a), third (b) and fourth (c) order (in
${\bar{r}}_0)$ connected diagrams for $\langle
Z^3\rangle\mid_{{\bar{v}}_0 =0}$.  In fourth order, there
are a few other similar diagrams which contribute to the
leading divergence.  For connectedness, the series has to
start at order two. (d) A few diagrams which contribute to
subleading divergences in the third and fourth order in
${\bar{r}}_0$\label{z3}}
\figure{Four chain diagrams for the annealed problem with
four chains.  The wiggly lines represent ${\bar{r}}_0$ type
interaction, and a solid horizontal line connecting three
chains is the three chain $\delta$-function interaction.
The first three terms of the series involving only four
chain interactions are shown in (a), (b), and (c). (d and
e) Two cases involving both the three and four chain
interactions. (f) A possible diagram in second order with
different chain combinations for the interactions.  This
contributes in the subleading divergence.\label{zann}}
\figure{Diagrams for $\langle Z\rangle$ upto second order
in $v_0$ and $v_0b(z)$.  The wavy and dotted lines
interactions with coupling constants $v_0 b(z)$ and $v_0$
respectively. \label{zav}}
\figure{(a) Diagrams involving only $v_0b(z)$ for $\langle
Z^2\rangle$. (b) Diagrams of (a) after disorder averaging.
(See the caption of Fig ~\ref{z2}.) Fig (a2) with odd
number of wavy lines vanishes after DA.  Different pairings
lead to two possiblities (b2) and (b3) from (a3).
Similarly for a4, there are two diagrams (b4) and (b5)
after DA.  Diagrams (b2) and (b4) are not considered for
the $\langle Z^2\rangle\mid_{{\bar{v}}_0 =0}$ case.
\label{z2av}}
\end{document}